\documentclass[aps,prl,twocolumn,superscriptaddress]{revtex4}
\usepackage{graphicx}
\usepackage{amsmath}
\usepackage{amssymb}
\usepackage{colordvi}
\usepackage{mathrsfs}
\usepackage{bm}
\usepackage{verbatim}
\usepackage{dcolumn}
\usepackage{bm}
\usepackage{epsfig}
\usepackage{subfigure}
\usepackage{siunitx}
\usepackage{ulem}

\begin{document}
\title{Metallic Network of Topological Domain Walls}
\author{Tao Hou $^\dagger$}
\affiliation{ICQD, Hefei National Laboratory for Physical Sciences at Microscale, CAS Key Laboratory of Strongly-Coupled Quantum Matter Physics, and Department of Physics University of Science and Technology of China, Hefei, Anhui 230026, China}
\author{Yafei Ren $^\dagger$}
\affiliation{ICQD, Hefei National Laboratory for Physical Sciences at Microscale, CAS Key Laboratory of Strongly-Coupled Quantum Matter Physics, and Department of Physics University of Science and Technology of China, Hefei, Anhui 230026, China}
\author{Yujie Quan}
\affiliation{ICQD, Hefei National Laboratory for Physical Sciences at Microscale, CAS Key Laboratory of Strongly-Coupled Quantum Matter Physics, and Department of Physics University of Science and Technology of China, Hefei, Anhui 230026, China}
\author{Jeil Jung}
\affiliation{Department of Physics, University of Seoul, Seoul 02504, South Korea}
\author{Wei Ren}
\affiliation{International Centre for Quantum and Molecular Structures, Materials Genome Institute, Shanghai Key Laboratory of High Temperature Superconductors, Physics Department, Shanghai University, Shanghai 200444, China}
\author{Zhenhua Qiao}
\email[Correspondence author:~]{qiao@ustc.edu.cn}
\affiliation{ICQD, Hefei National Laboratory for Physical Sciences at Microscale, CAS Key Laboratory of Strongly-Coupled Quantum Matter Physics, and Department of Physics University of Science and Technology of China, Hefei, Anhui 230026, China}
\date{\today}

\begin{abstract}
We study the electronic and transport properties of a network of domain walls between insulating domains with opposite valley Chern numbers.
%
We find that the network is semi-metallic with Dirac dispersion near the charge neutrality point and the corresponding electronic states distribute along the domain walls. Near the charge neutrality point, we find quantized conductance in nanoribbon with sawtooth domain wall edges that propagates along the boundaries and is robust against weak disorder. For a trident edged ribbon, we find a small energy gap due to the finite size effect making the nanoribbon an insulator.
When the Fermi energy is away from charge neutrality point, all domain walls contribute to the conduction of current.
Our results provide a comprehensive analysis of the electronic transport properties in a topological domain wall
network that not only agrees qualitatively with experiments on marginally twisted bilayer graphene under a perpendicular electric field, but also can provide useful insights for designing low-power topological quantum devices.
\end{abstract}
	
\maketitle
	
\textit{Introduction---.} Topological zero line modes (ZLMs), which are localized at the domain walls between domains with different valley Hall topologies, have been studied in gapped graphene systems~\cite{zlm1, QAH1, QSH, QAH2, zlm, anlysis, sciencejun,folded, HJ, partition, kink, 2013, 2015, 2016, NP, N2015, XY,Zhangfan,Naturepartition}.
A single domain wall and the intersection between two domain walls have been experimentally realized in bilayer graphene under precisely controlled electrical field that varies spatially~\cite{nature nano,Natphys,Nature}. Nevertheless, it is technologically challenging to devise a periodic network of intersecting topological domain walls using similar techniques~\cite{sciencejun}.
Fortunately, recent experiments on twisted layered materials provide a natural platform to explore such network~\cite{networkSTM2018, networkscience2018, Geim, networkAline2018, prbnetwork2018, networls-multi, network2013, science12, nethelin, networkwk, magneticfieldwk, PNAS, Kim}.

The twisted bilayer graphene around magic angle has been extensively explored due to the discovery of superconductivity, topological phases and correlated insulating states~\cite{caoyuan1,Nandkishore,J,M,T,L}. Interesting physics at other twisted angles are also reported~\cite{Kim}. When the twist angle decreases from the magic angle of $1.1^{\circ}$, the incommensurate moir\'{e} structure gradually becomes an array of commensurate domains with soliton boundaries. Around $0.1^{\circ}$, periodic AB/BA domains with sharp domain walls are much similar to the previous reports on AB/BA stacking graphene bilayer formed in chemically vapour-deposited ~\cite{Kim}~\cite{Kim}. When a perpendicular electric field is applied via electric gating, AB and BA domains become insulating with opposite valley Chern numbers while the ZLMs along domain walls form a conducting network~\cite{folded, partition, kink, 2013, 2015, 2016, NP, N2015}.
These networks have been imaged in minimally twisted graphene bilayers through scanning tunneling microscopy~\cite{networkSTM2018, Geim, nethelin, science12} and optical techniques~\cite{science12}.
Similar network of domain wall states could also arise in graphene/hexagonal boron nitride heterostructures~\cite{pressure, GrBN}.
However, the electronic transport properties through domain wall networks remain poorly understood~\cite{networkSTM2018, networkscience2018, Geim, networkAline2018, prbnetwork2018, networls-multi, network2013, science12, nethelin, networkwk}.

\begin{figure*}
  \includegraphics[width=13cm,angle=0]{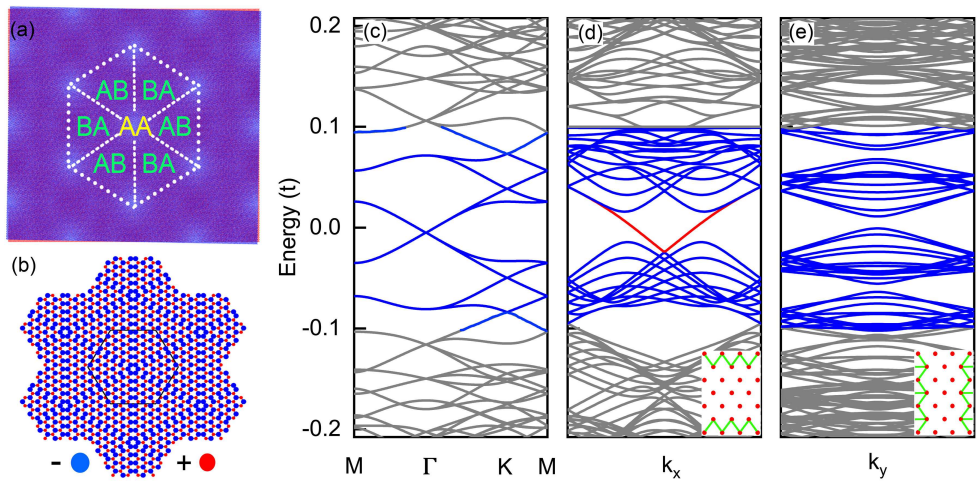}
  \caption{(a) Schematic plot of twisted bilayer graphene. AB/BA/AA stacking regions are marked. (b) Monolayer graphene with position-dependent lattice potentials. The magnitude of sublattice potential is set to be $\varDelta=0.1t$ with positive and negative signs on sites shown in red and blue, separately. (c) Bulk band structure along high-symmetry lines in the Brillouin zone of the domain wall network. (d)-(e) Band structures of ribbons illustrated in the insets with sawtooth and trident boundaries, separately. The side length of the bulk primitive cell is 8.1 nm and the width of the ribbon unit cell is 24 nm. Blue bands are inside the bulk band gap, and the red bands highlight the gapless edge modes propagating along the boundaries.}
  \label{fig1}
\end{figure*}
	
In this Letter, we focus on the electronic and transport properties of a domain-wall network formed by periodically arranged domains with opposite valley Chern numbers of $\pm 1$, which is modeled through graphene lattices with staggered potentials. The domain walls intersect with each other and the intersecting points form a triangular superlattice. Using tight-binding model, we calculate the electronic structures and find that though each topological domain is insulating with a bulk gap, the ZLMs that propagate along the network of domain walls are metallic with linear Dirac dispersion.
In a nanoribbon with trident edges a small energy gap is opened due to finite-size effect, whereas in a ribbon with sawtooth edges we find nearly quantized conductance $G=e^2/h$ through
gapless modes distributed at system boundaries that are robust against weak disorder.
Similar gapless modes and finite conductivity are found for arbitrary boundaries deviate from a trident edge configuration,
indicating that the latter is a special case.
However, when the Fermi energy deviates away from charge neutrality point (CNP), the transport
becomes independent of the system's edge configuration as the current spreads along all domain walls in the network.
The quantization of transport near CNP is in qualitative agreement with recent transport experiments
on a network of domain walls in twisted bilayer graphene under strong electric field~\cite{Geim, Kim}.

\textit{System model and gapless energy bands---.} A twisted bilayer graphene is shown schematically in Fig.~\ref{fig1}(a), where the bright zone in the center corresponds to AA stacking. Around this central zone, AB/BA stacking domains are formed periodically. By applying a perpendicular electric field, the AB/BA stacking domains become gapped with opposite valley Chern numbers, while the domain walls between AB and BA stacking regions exhibit gapless ZLMs~\cite{networkSTM2018, nethelin, science12}.
Without loss of generality, we consider a network of domain walls in a simpler model in monolayer graphene with spatially varying staggered sublattice potentials as illustrated in Fig.~\ref{fig1}(b) where sites in red/blue have positive/negative site energies.
These geometries have the topological domain wall structure of Fig.~\ref{fig1}(a),
and can be described by the following tight-binding Hamiltonian:
\begin{eqnarray}
H=-t \sum_{ \langle ij \rangle} c_i^{\dag} c_j+\sum_{ i\in A}U_{A}c_i^{\dag}c_i+\sum_{ j\in B}U_{B}c_j^{\dag}c_j, \nonumber
\end{eqnarray}
where $c_i^{\dag}(c_i)$ is a creation (annihilation) operator for an electron at site $i$, and $t=2.6$~eV is the nearest-neighbor hopping amplitude.
The sublattice potentials are $U_{A}$=$-U_{B}$=$\lambda\varDelta$ with $\lambda = \pm 1$ in the AB/BA stacking regions, where 2$\varDelta$ measures the magnitude of energy gaps at those domains.
The lattice distance is set to $a=0.14~$nm in our calculations.	
The unit cell of superlattice is indicated with a black hexagon in Fig.~\ref{fig1}(b) and
the bulk band structure is plotted along the high-symmetry lines in Fig.~\ref{fig1}(c) when
the system has a bulk band gap $ 2 \varDelta=0.2t$.
We can observe in-gap band structures within the bulk gap consisting of Dirac-like
linear bands around $\Gamma$ point, and additional mini Dirac cones at higher energies around $K/K'$ points
interspersed by regions with narrow bandwidths.
Since the numbers of sites with positive and negative on-site potential are different, the resulting nonzero averaged value leads to the shift of Dirac point.

We further calculate the ribbon band structures with sawtooth and trident boundary conditions as shown in Figs.~\ref{fig1}(d) and \ref{fig1}(e), respectively.
In the former case, we find gapless states inside the bulk band gap as shown in blue. Near the charge neutrality point, bands with linear dispersion appear as plotted in red that propagate along boundaries as shown later. In the later case, a small avoided gap appears at CNP due to finite-size effect.
These features indicate that a triangular ZLMs network in the bulk will behave as a Dirac metal
at CNP while the band structure of nanoribbons depends on the boundary conditions.

\begin{figure}
	\includegraphics[width=7 cm,angle=0]{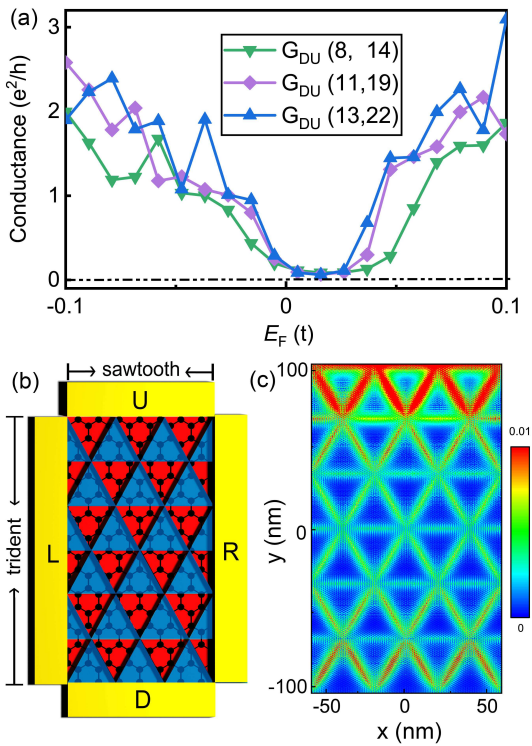}
		\caption{ (a) Dependence of conductance as a function of Fermi level from lead U to D for different system sizes ($L_x, L_y$).
			(b) Schematic of four-terminal network device formed by nine intersecting nodes with length of ($L_x, L_y$). The sample shows sawtooth (trident) boundary condition from leads L to R (U to D). Red and blue regions correspond to the gapped domains with opposite topologies.
			(c) Local density of stats of current injected from U to D with a trident boundary at $E_{\rm F}=0.050t$. We set $L_x=$110 nm and $L_y=220$ nm for (c).
		    }
\label{fig2}
\end{figure}

\textit{Transport properties.---} We consider the transport properties through a rectangular nano-flake of domain wall network as shown in  Fig.~\ref{fig2}(b) where four leads are connected, labeled by R, L, U, and D.
The transport calculations are performed by employing Landauer-B\"{u}ttiker
formula~\cite{Datta} and recursively constructed Green's functions~\cite{wang}.
The conductance from lead $q$ to $p$ is evaluated by $G_{pq}=\frac{2e^2}{h} {\rm Tr}[\Gamma_{p} G^r \Gamma_{q} G^a]$ where $G^{r,a}$ is the retarded/advanced Green's function of the central scattering region, and $\Gamma_{p}$ is the line-width function describing the coupling between lead $p$ and the central scattering region. The propagation of currents injected from lead $p$ at energy $\epsilon$ is illustrated by the local density of states (LDOS) $\rho_p(r,\epsilon)=1/2\pi[G^r\Gamma_{p}G^a]_{rr} $ where $r$ is the spatial coordinate.
The sample shows sawtooth edge from leads L to R and trident edge from U to D as illustrated in Fig.~\ref{fig2}(b).
In our calculation, we use staggered potentials of $\Delta=0.1t$ and take a sample of width $L_x=59$~nm and length $L_y=102$~nm.

We first calculate the two-terminal conductances from U to D
with trident edges for different Fermi energies and system sizes as shown in Fig.~\ref{fig2}(a).
In this vertical UD ribbon geometry setup we find a band gap which is signaled vanishing current flowing into lead D near charge neutrality point and a rapid increase of conductance for finite carrier doping, as show in Fig.~\ref{fig2}(c) from the plots of the LDOS of currents injected from U at the Fermi energy of $E_{\rm F}=0.05t$.
The gap size at CNP is found to decrease with increasing network size.

\begin{figure}
	\includegraphics[width=7 cm,angle=0]{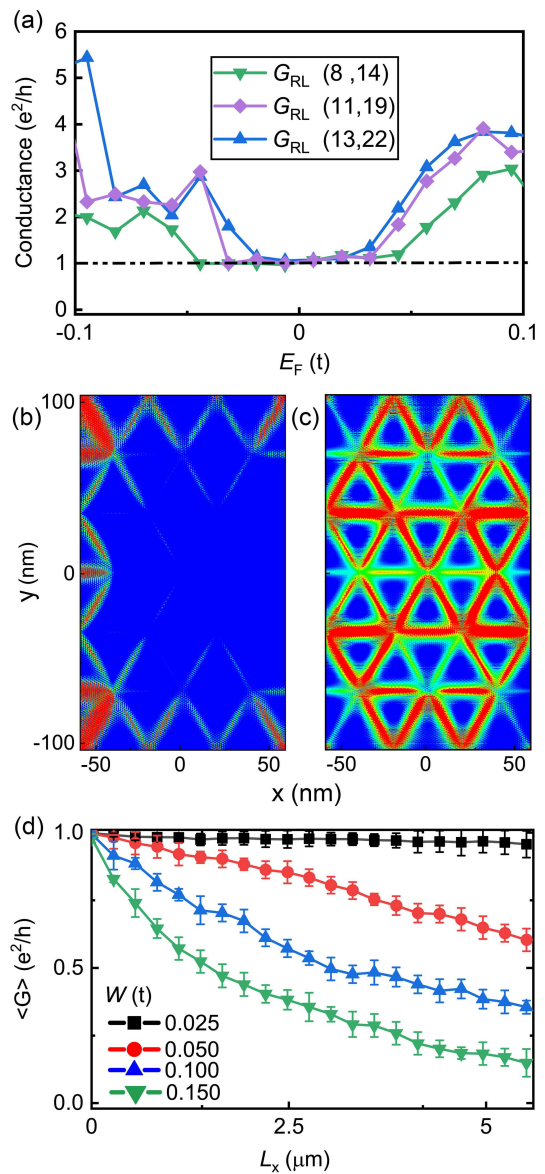}
	\caption{(a) Dependence of conductance as a function of Fermi level from lead L to R for different system sizes ($L_x, L_y$).
		(b)-(c) LDOS of current injected from L to R with sawtooth domain wall boundary at (b) $E_{\rm F}=0.001t$ and (c) for $E_{\rm F}=0.050t$. We set $L_x=$110 nm and $L_y=220$ nm for (b) and (c).
	    (d) Averaged conductance $\langle G_{LR} \rangle$ from L to R as function of the system length $L_x$ for various disorder strengths at $L_y = 22$ nm when the Fermi energy is set to be at CNP.
	    Over 50 samples are collected for each data point. The unit of the disorder strength is $t$. }
	\label{fig3}
\end{figure}
\begin{figure*}
	\includegraphics[width=13 cm,angle=0]{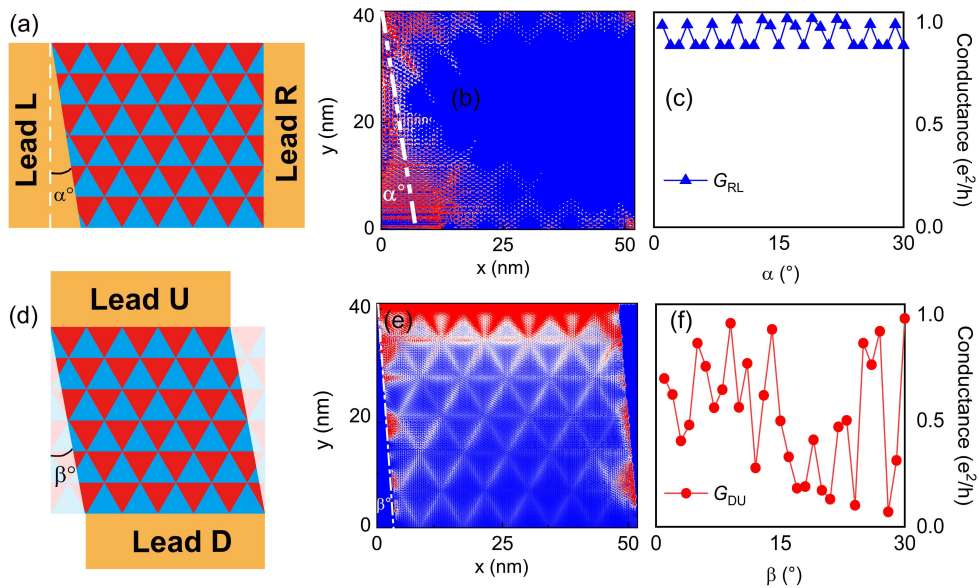}
	\caption{(a) Schematic of two-terminal network device along the sawtooth edge with tilt angle $\alpha$ between lead L and sample.
	 (b) LDOS for current injecting from L to R at $\alpha=\ang{10}$.
	 (c) Dependence of conductance on tilt angles $\alpha$.
	 (d) Schematic of two-terminal network device from U to D with a rough edge instead of the trident edge, the included angle between them is $\beta$.
	 (e) LDOS for current injecting from U to D at $\beta=\ang{5}$.
	 (f) Dependence of conductance on included angle $\beta$.}	
	\label{fig4}
\end{figure*}

Then we calculate the conductance between L and R for different Fermi energies and system sizes as shown Fig.~\ref{fig3}(a). We find that the conductance $G_{LR}$ remains quantized at $e^2/h$ around CNP.
As the size of the network grows, the energy window of quantized $G_{LR}$ shrinks. The LDOS for current injected from lead L at Fermi energies near and away from CNP are plotted in Figs.~\ref{fig3}(b) and \ref{fig3}(c), respectively. Near the CNP, we find that the incoming current propagates along the sawtooth shaped domain walls near the upper and lower boundaries. It is noteworthy that the total conductance contributed from both sides is close to a conductance quantum, i.e., $e^2/h$, which is attributed to the edge modes shown in red in Fig.~\ref{fig1}(d). The corresponding edge state wavefunction is distributed at both sides of the sample even though the two boundaries are well separated.
Based on our numerical results on the monolayer honeycomb lattice with valley Hall domains, we expect that the conductance of a zero-line network in twisted bilayer graphene consisting of two layers and spin degree of freedom should be quadruple $e^2/h$, which agrees with recent experimental observations~\cite{Kim}.
When the Fermi energy is shifted away from CNP, the incoming current is partitioned in the whole topological network as shown in Fig.~\ref{fig3}(c).
Thus, away from CNP the anisotropy of transport is negligibly small.

To further show the robustness of the transport properties against random disorder near
CNP, we investigate the effect of Anderson disorder in the sample bulk. 
The Anderson disorder is introduced through  random on-site potentials ranging between $[-W/2, W/2]$ where $W$ characterizes the disorder strength.
We calculate the conductance at CNP of two-terminal devices with sawtooth boundary condition for different $W$ and length $L_x$ ranging from several nanometers to micrometers as shown in Fig.~\ref{fig3}(d) by keeping $L_y = 22~$nm.
From this figure, we find that for small disorder strength (e.g., $W = 0.025t$), the edge modes are nearly ballistic since the quantization of the conductance remains very robust and shows a weak dependence on $L_x$.
For stronger disorder strengths, the conductances remain robust. For example, at a disorder magnitude of $W=0.1t$ that is comparable to the bulk band gap, the conductance is still half of $e^2/h$ for a sample with a length of $2.5~\mu$m.

\textit{Role of boundary conduction---.}
The quantization of the conductance remains when we introduce disorder between the contact leads and samples by changing the tilt angle $\alpha$ between lead and sample as illustrated in Fig.~\ref{fig4}(a). We plot the conductance in Fig.~\ref{fig4}(c) by a blue line with triangles, where we find that the conductance is nearly quantized and shows a weak dependence on $\alpha$.

The electrical conductance $G_{DU}$ of a trident edged ribbon vanishes at CNP but it becomes nonzero as soon as the edge geometry varies slightly.
We consider a ribbon with rough edges as shown in Fig.~\ref{fig4}(d) where the atoms in regions
with angle $\beta$ at both sides are removed.
At $\beta = 5^\circ$, the LDOS is plotted in Fig.~\ref{fig4}(e) where one can find that finite current propagates along the sample boundary.
In Fig.~\ref{fig4}(f), we plot the dependence of conductance $G_{DU}$ on $\beta$.
We find that, though the conductance is not quantized, $G_{DU}$ at CNP becomes nonzero when the boundary deviates from the trident edge.

\textit{Summary---.} We presented a systematic study of the electronic and transport properties of domain wall networks formed by periodically arranged insulating domains with opposite valley Chern numbers.
We find that the network gives rise to a metallic electronic structure with Dirac dispersions near the CNP. In a ribbon with sawtooth boundary condition, gapless edge modes are found with the corresponding wave-function distributed at both sides of the edge that are well separated in space.
The edge modes contribute to one conductance quantum $e^2/h$ with current flowing along both sides. By including Anderson type disorder, we find that the quantization of conductance at CNP is quite robust, and is weakly dependent on the details of the contact between the metallic lead and the central region.
We also find that, although the conductance vanishes at CNP for a ribbon with
trident boundary conditions, the system becomes conducting when the edge geometry varies,
in agreement with the finite conductance observed in experiments.
If we generalize the results obtained in monolayer graphene system to a twisted bilayer and include the spin degree of freedom,
we expect a quantized conductance of $4e^2/h$ near CNP,
in keeping with the recent experimental observations for the transport in marginally twisted bilayer
graphene under the effects of an electric field.

Our theoretical proposal can find its experimental realization in moir\'e structures of graphene/h-BN heterostructure, in twisted bilayer graphene with a perpendicular electric field, or phononic crystals.
Specifically, our work reports the first microscopic prediction of the transport in a network of triangular topological channels in minimally twisted bilayer graphene,
and paves the way for understanding the transport properties in other triangular domain wall network
superlattice systems made through graphene multilayers and other 2D materials.

\begin{acknowledgments}
\textit{Acknowledgments---.} This work was supported financially by the National Key R \& D Program (2017YFB0405703), and the NNSFC (11474265, and 11674024), the China Government Youth 1000-Plan Talent Program, and Anhui Initiative in Quantum Information Technologies. We are grateful to AMHPC and Supercomputing Center of USTC for providing high-performance computing assistance. J. J. acknowledges financial support from the Samsung Science and Technology Foundation under project no. SSTF-BA1802-06.
\end{acknowledgments}

$^\dagger$ These authors contributed equally to this work.

\end{document}